\documentstyle[12pt,epsf,rotate]{article}
\input{psfig}

\newcommand{\gsim}{~{}_{\textstyle\sim}^{\textstyle >}~}

\def\OEE{\Omega_{\eta+\eta'}}

\newcommand{\RE}{{\rm Re}}
\newcommand{\IM}{{\rm Im}}
\newcommand{\vcb}{|V_{cb}|}

\newcommand{\vub}{|V_{ub}|}

\newcommand{\vus}{|V_{us}|}

\def\eps{\varepsilon}
\def\epe{\varepsilon'/\varepsilon}

\newcommand{\mt}{m_{\rm t}}

\newcommand{\mc}{m_{\rm c}}
\newcommand{\ms}{m_{\rm s}}
\newcommand{\md}{m_{\rm d}}

\newcommand{\mw}{M_{\rm W}}
\newcommand{\mz}{M_{\rm Z}}
\newcommand{\gev}{\, {\rm GeV}}
\newcommand{\mev}{\, {\rm MeV}}
\newcommand{\bsi}{B_6^{(1/2)}}
\newcommand{\bei}{B_8^{(3/2)}}
\newcommand{\Lms}{\Lambda_{\overline{\rm MS}}}

\newcommand{\bea}{\begin{eqnarray}}
\newcommand{\eea}{\end{eqnarray}}
\newcommand{\bd}{\begin{displaymath}}
\newcommand{\ed}{\end{displaymath}}

\newcommand{\beq}{\begin{equation}}
\newcommand{\eeq}{\end{equation}}
\newcommand{\be}{\begin{equation}}
\newcommand{\ee}{\end{equation}}
\newcommand{\bi}{\begin{itemize}}
\newcommand{\ei}{\end{itemize}}
\newcommand{\ord}{{\cal O}}

\begin{document}
\thispagestyle{empty}
\phantom{xxx}
\vskip1truecm
\begin{flushright}
 TUM-HEP-355/99 \\
August 1999
\end{flushright}
\vskip1.8truecm
\centerline{\LARGE\bf Theoretical Status of $\epe$}
   \vskip1truecm
\centerline{\Large\bf Andrzej J. Buras}
\bigskip
\centerline{\sl Technische Universit{\"a}t M{\"u}nchen}
\centerline{\sl Physik Department} 
\centerline{\sl D-85747 Garching, Germany}
\vskip1truecm

\phantom{xxx} \vspace{-6mm}

\begin{abstract}
We review the present theoretical status of the CP violating ratio 
$\epe$ in the Standard Model (SM) and confront its estimates with the
most recent data.
In particular we review the present status of the
most important parameters $\ms$, $\bsi$, $\bei$, 
$\Lms^{(4)}$ and $\OEE$.
While the sign and the order of magnitude of SM
estimates for $\epe$ agree with the data,
for  central values of these parameters most estimates of
 $\epe$ in the SM are substantially below
the grand experimental average $ (21.2\pm 4.6)\cdot 10^{-4}$.
Only in a small corner of the parameter space
can $\epe$ in SM  be made
consistent with experimental results. 
In view of very large theoretical 
uncertainties, it 
is impossible to conclude at present that the data on $\epe$ indicate
new physics.
A brief discussion of $\epe$ beyond the Standard Model
is presented.
\end{abstract}

\vskip1truecm

\centerline{\it Invited talk given at }
\centerline{\bf KAON 99}
\centerline{\sl Chicago, June 21-26, 1999}
%%% end title page %%%%%%%%%%%%%

\newpage

\setcounter{page}{1}

\section{Introduction}
The purpose of this talk is to summarize the present theoretical
status of the CP violating ratio $\epe$ and to confront it with
the recent experimental findings. The parameters $\varepsilon$ and
$\varepsilon'$ describe two types of CP violation in the
decays $K_L\to \pi\pi$, which could not take place if CP
was conserved. In the Standard Model a non-vanishing value
of $\varepsilon$ originates in the fact that the mass eigenstate
$K_L$ is not a CP eigenstates due to the complex CKM couplings
\cite{KM}
in the box diagrams responsible for the $K^0-\bar K^0$ mixing.
Indeed, $K_L$ is a linear combination
of CP eigenstates $K_2$ (CP=$-)$ and $K_1$ ( CP=$+$):
$K_L=K_2+\bar\varepsilon K_1$, where $\bar\varepsilon$ is
a small parameter. The decay of
$K_L$ via $K_1$ into CP=$+$ state $\pi\pi$ is termed {\it indirectly}
CP violating as it
proceeds not via explicit breaking of the CP symmetry in 
the decay itself but via the admixture of the CP state with opposite 
CP parity to the dominant one.
 The measure for this
indirect CP violation is defined as
\begin{equation}\label{ek}
\varepsilon
={{A(K_{\rm L}\rightarrow(\pi\pi)_{I=0}})\over{A(K_{\rm 
S}\rightarrow(\pi\pi)_{I=0})}},
\end{equation}
which can be rewritten as
\begin{equation}
\eps = \frac{\exp(i \pi/4)}{\sqrt{2} \Delta M_K} \,
\left( \IM M_{12} + 2 \xi \RE M_{12} \right),
\quad\quad
\xi = \frac{\IM A_0}{\RE A_0}~,
\label{eq:epsdef}
\end{equation}
where $ M_{12}$ represents $K^0-\bar K^0$ mixing, $\Delta M_K$ is
the $K_L-K_S$ mass difference and $A_0$ is the isospin amplitude
for $\pi\pi$ in the $I=0$ state. The second term
involving $\xi$ cancells 
the phase convention dependence of the first term but in the usual
CKM convention can be safely neglected. In this limit 
$\varepsilon=\bar\varepsilon$.

On the other hand, so-called {\it direct}
CP violation is realized via a 
direct transition of a CP odd to a CP even state or vice versa.
$K_2\to\pi\pi$ in the case of $K_L\to\pi\pi$.
A measure of such a direct CP violation in $K_L\to \pi\pi$ is characterized
by a complex parameter $\varepsilon'$  defined as
\be\label{eprime}
\varepsilon'=\frac{1}{\sqrt{2}}\IM\left(\frac{A_2}{A_0}\right)
              \exp(i\Phi_{\varepsilon'}), 
   \qquad \Phi_{\varepsilon'}=\frac{\pi}{2}+\delta_2-\delta_0~.
\ee
Here the subscript $I=0,2$ denotes $\pi\pi$ states with isospin $0,2$
equivalent to $\Delta I=1/2$ and $\Delta I = 3/2$ transitions,
respectively, and $\delta_{0,2}$ are the corresponding strong phases. 
The weak CKM phases are contained in $A_0$ and $A_2$.
The isospin amplitudes $A_I$ are complex quantities which depend on
phase conventions. On the other hand, $\varepsilon'$ measures the 
difference between the phases of $A_2$ and $A_0$ and is a physical
quantity.
The strong phases $\delta_{0,2}$ can be extracted from $\pi\pi$ scattering. 
Then $\Phi_{\varepsilon'}\approx \pi/4$.

Experimentally we have \cite{CRONIN,PDG}
\begin{equation}\label{eexp}
\varepsilon_{\rm exp}
=(2.280\pm0.013)\cdot10^{-3}\;\exp{i\Phi_{\varepsilon}},
\qquad \Phi_{\varepsilon}\approx{\pi\over 4}.
\end{equation}
Until recently the experimental situation on $\epe$ was rather unclear:
\begin{equation}\label{eprime2}
\RE(\varepsilon'/\varepsilon) =\left\{ \begin{array}{ll}
(23 \pm 7)\cdot 10^{-4} & {\rm (NA31)}~\cite{barr:93} \\
(7.4 \pm 5.9)\cdot 10^{-4} &{\rm (E731)}~ \cite{gibbons:93}.
\end{array} \right.
\end{equation}
While the result of the NA31 collaboration at CERN 
\cite{barr:93}
clearly indicated direct CP violation, the value of E731 at Fermilab
\cite{gibbons:93}, was compatible with superweak theories
\cite{wolfenstein:64} in which $\varepsilon'/\varepsilon = 0$.
After heroic efforts on both sides of the Atlantic,
this controversy is now settled with the two new measurements
by KTeV at Fermilab \cite{KTEV} and NA48 at CERN \cite{NA48}
\begin{equation}\label{eprime1}
\RE(\varepsilon'/\varepsilon) =\left\{ \begin{array}{ll}
(28.0 \pm 4.1)\cdot 10^{-4} & {\rm (KTeV)}~\cite{KTEV} \\
(18.5 \pm 7.3)\cdot 10^{-4} &{\rm (NA48)}~ \cite{NA48},
\end{array} \right.
\end{equation}
which together with the NA31 result confidently establish direct 
CP violation in nature ruling out superweak models \cite{wolfenstein:64}.
The grand average including NA31, E731, KTeV and NA48 results reads
\cite{NA48}
\be
\RE(\epe) = (21.2\pm 4.6)\cdot 10^{-4}
\label{ga}
\ee
very close to the NA31 result but with a smaller error. 
The error 
should be further reduced once 
complete data from both collaborations have 
been analyzed. It is also 
of great interest to see what value for $\epe$ will be measured by KLOE at 
Frascati, which uses a different experimental technique than KTeV and NA48.

\section{$\eps$ in the Standard Model}\label{Epsilon}
\setcounter{equation}{0}
In the Standard Model $\eps$ is found by calculating the box diagrams
with internal $u,~c,~t,~W^\pm $ exchanges and including short distance
QCD corrections. The final result can be written as follows
\begin{equation}
\eps=C_{\eps} \hat B_K \IM\lambda_t \left\{
\RE\lambda_c \left[ \eta_1 S_0(x_c) - \eta_3 S_0(x_c, x_t) \right] -
\RE\lambda_t \eta_2 S_0(x_t) \right\} \exp(i \pi/4)\,,
\label{eq:epsformula}
\end{equation}
where $C_{\eps}$ is a numerical factor, $\lambda_i=V^*_{is}V_{id}$
and $S_0(x_i)$ with $x_i=m^2_i/\mw^2$ are Inami-Lim functions
\cite{IL}. Explicit expressions can be found in \cite{AJBLH}. 
The NLO values of the QCD factors $\eta_i$ are given as follows 
\cite{NLOS}:
\begin{equation}
\eta_1=1.38\pm 0.20,\qquad
\eta_2=0.57\pm 0.01,\qquad
  \eta_3=0.47\pm0.04~.
\end{equation}
The main theoretical uncertainty in (\ref{eq:epsformula})
resides in the non-perturbative parameter $\hat B_K$ which
parametrizes the relevant hadronic matrix element
$\langle \bar K^0| (\bar s d)_{V-A} (\bar s d)_{V-A} |K^0\rangle$.
Recent reviews of $\hat B_K$ are given in \cite{GUPTA98,EP99,AJBLH}.
In our numerical analysis presented 
below we will use 
$\hat B_K=0.80\pm 0.15$
which is in the ball park of various lattice and large-N estimates.

It is well known that that the experimental value of $\eps$ in
(\ref{eexp}) can be accomodated in the Standard Model. We know
this from numerous analyses of the unitarity triangle which
in addition to (\ref{eexp}) take into account data on 
$B^0_{d,s}-\bar B^0_{d,s}$ mixings and the values of the
CKM elements $\vus$, $\vub$ and $\vcb$. From this analysis
one extracts in particular \cite{EP99}
\begin{equation}\label{imt}
\IM\lambda_t =\left\{ \begin{array}{ll}
(1.33 \pm 0.14)\cdot 10^{-4} & {\rm (Monte~Carlo)} \\
(1.33 \pm 0.30)\cdot 10^{-4} &{\rm (Scanning)}.
\end{array} \right.
\end{equation}
 The ``Monte Carlo" error stands for an analysis in which the
experimentally measured parameters, like $\vus$, $\vub$ and $\vcb$,
are used with Gaussian errors and the theoretical input parameters,
like $\hat B_K$, $F_B\sqrt{\hat B_B}$ are taken with flat
distributions. In the ``scanning" estimate all input parameters
are scanned independently in the appropriate ranges. Details
can be found in \cite{EP99}. This analysis gives also
$\sin 2\beta=0.73\pm0.09~(0.71\pm0.13)$, where $\beta$ is one
of the angles in the unitarity triangle. The recent study
of CP violation in $B^0\to J/\psi K_S$ by CDF \cite{CDF99} gives
$\sin 2\beta=0.79\pm0.44$ in good agreement with the value above,
although the large experimental error precludes any definite
conclusion.

%\clearpage
\section{$\epe$ in the Standard Model}\label{EpsilonPrime}
\setcounter{equation}{0}
\subsection{Basic Formulae}
In the Standard Model the ratio $\epe$ is governed by
QCD
penguins and electroweak penguins ($\gamma$ and $Z^0$ penguins).
With increasing value of $\mt$ the electroweak penguins become
increasingly important \cite{FL,BU} and entering $\epe$ with the opposite
sign to QCD penguins, suppress this ratio for large $\mt$.
The size of $\epe$ is also strongly affected by 
QCD renormalization group effects. Without these effects $\epe$
would be at most $\ord(10^{-5})$ independently of $\mt$ \cite{BU}.

The parameter $\varepsilon'$ is given in terms of the isospin amplitudes
$A_I$ in (\ref{eprime}). Applying the operator product expansion
 to these amplitudes one finds
\begin{equation}
\frac{\varepsilon'}{\varepsilon} = 
\IM \lambda_t\cdot F_{\varepsilon'}(\mt,\Lms^{(4)},\ms,\bsi,\bei,\OEE),
\label{eq:epe1}
\end{equation}
where
\begin{equation}
F_{\varepsilon'} = 
\left[ P^{(1/2)} - P^{(3/2)} \right] \exp(i\Phi),
\qquad \Phi=\Phi_{\varepsilon'}-\Phi_\varepsilon 
\label{eq:epe2}
\end{equation}
with
\begin{eqnarray}
P^{(1/2)} & = & r \sum y_i(\mu) \langle Q_i(\mu)\rangle_0
(1-\Omega_{\eta+\eta'}) ~,
\label{eq:P12} \\
P^{(3/2)} & = &\frac{r}{\omega}
\sum y_i(\mu) \langle Q_i(\mu)\rangle_2~.~~~~~~
\label{eq:P32}
\end{eqnarray}
Here
\begin{equation}
r = \frac{G_{\rm F} \omega}{2 |\eps| \RE A_0}~, 
\qquad
\langle Q_i\rangle_I \equiv \langle (\pi\pi)_I | Q_i | K \rangle~,
\qquad
\omega = \frac{\RE A_2}{\RE A_0}
\label{eq:repe}
\end{equation}
and $\mu$ is the renormalization scale which is $\ord(1\gev)$.
Since
$\Phi \approx 0$,
$F_{\varepsilon'}$ and $\epe$
are real  to an excellent approximation. The arguments of
$F_{\varepsilon'}$ will be discussed shortly.
Explicit expressions for the operators $Q_{1,2}$ (current-current),
$Q_{3-6}$ (QCD-penguins) and $Q_{7-10}$ (electroweak penguins)
are given in \cite{AJBLH}. The dominat are these two:
\begin{equation}\label{OS3}
 Q_6 = (\bar s_{\alpha} d_{\beta})_{V-A}\sum_{q=u,d,s}
       (\bar q_{\beta} q_{\alpha})_{V+A},
\qquad
 Q_8 = {3\over2}\;(\bar s_{\alpha} d_{\beta})_{V-A}\sum_{q=u,d,s}e_q
        (\bar q_{\beta} q_{\alpha})_{V+A}
\end{equation}
where, $e_q$ denotes the electric quark charges.

The Wilson coefficient functions $ y_i(\mu)$
were calculated including
the complete next-to-leading order (NLO) corrections in
\cite{NLOP}--\cite{ROMA1}. The details
of these calculations can be found there and in the review
\cite{BBL}. 
Their numerical values for $\Lms^{(4)}$ corresponding to
$\alpha_{\overline{MS}}^{(5)}(\mz)=0.119\pm 0.003$
and two renormalization schemes (NDR and HV)
can be found  in  \cite{EP99}.

It is customary in phenomenological
applications to take $\RE A_0$ and $\omega$ from
experiment, i.e.
\begin{equation}
\RE A_0 = 3.33 \cdot 10^{-7}\gev,
\qquad
\omega = 0.045,
\label{eq:ReA0data}
\end{equation}
where the last relation reflects the so-called $\Delta I=1/2$ rule.
This strategy avoids to a large extent the hadronic uncertainties 
in the real parts of the isospin amplitudes $A_I$.

The sum in (\ref{eq:P12}) and (\ref{eq:P32}) runs over all contributing
operators. $P^{(3/2)}$ is fully dominated by electroweak penguin
contributions. $P^{(1/2)}$ on the other hand is governed by QCD penguin
contributions which are suppressed by isospin breaking in the quark
masses ($m_u \not= m_d$). The latter effect is described by
\cite{donoghueetal:86}-\cite{lusignoli:89} 
\begin{equation}
\Omega_{\eta+\eta'} = \frac{1}{\omega} \frac{(\IM A_2)_{\rm
I.B.}}{\IM A_0}=0.25 \pm 0.08\,.
\label{eq:Omegaeta}
\end{equation}

\subsection{History of $\epe$}
There is a long history of calculations of $\epe$ in the Standard
Model. As it has been already described in \cite{AJBLH,EP99},
I will be very brief here.
The first calculation of $\epe$ for $\mt \ll \mw$ without the inclusion
of renormalization group effects can be found in \cite{EGN}. 
Renormalization group effects
in the leading
logarithmic approximation have been first presented in \cite{GW79}. 
For $\mt \ll \mw$ only QCD
penguins play a substantial role. 
With increasing $\mt$ the role of electroweak penguins becomes 
important. The first leading log analyses for arbitrary $\mt$
can be found in
\cite{FL,BU}, where 
a strong cancellation between QCD penguins
and electroweak penguin contributions to $\epe$ for $m_t > 150~\gev$ 
has been found.
Finally,
during the nineties considerable progress has been made by
calculating complete NLO corrections to the Wilson coefficients
relevant for $\varepsilon$ \cite{NLOS} and $\varepsilon'$
\cite{NLOP}--\cite{ROMA1}. The progress in calculating the
corresponding hadronic matrix elements was substantially slower.
It will be summarized below. 

Now, the function $F_{\varepsilon'}$ in (\ref{eq:epe1}) can be written
in a crude approximation (not to be used for any serious analysis)
as follows
\begin{eqnarray}\label{ap}
F_{\varepsilon'}
&\approx& 13\cdot 
\left[\frac{110\mev}{\ms(2~\gev)}\right]^2
\left(\frac{\Lms^{(4)}}{340~\mev}\right)
\nonumber\\
& & \cdot
\left[\bsi(1-\OEE)-0.4\cdot \bei\left(\frac{\mt}{165\gev}\right)^{2.5}\right]
~.
\end{eqnarray}

Here $B_i$ 
are hadronic parameters defined through 
\begin{equation}
\langle Q_6 \rangle_0 \equiv B_6^{(1/2)} \, \langle Q_6
\rangle_0^{\rm (vac)}\,,
\qquad
\langle Q_8\rangle_2 \equiv B_8^{(3/2)} \, \langle Q_8
\rangle_2^{\rm (vac)} \,.
\label{eq:1}
\end{equation}
The label ``vac'' stands for the vacuum
insertion estimate of the hadronic matrix elements in question 
for
which $B_6^{(1/2)}=B_8^{(3/2)}=1$.
The dependence on $m_s$ in (\ref{ap}) originates in the $m_s$
dependence of $\langle Q_6\rangle_0^{\rm (vac)}$ and
$\langle Q_8\rangle_2^{\rm (vac)}$, so that
$\langle Q_6\rangle_0$ and
$\langle Q_8\rangle_2$
are roughly proportional
to
\be\label{RS}
R_6\equiv \bsi\left[ \frac{137\mev}{\ms(\mc)+\md(\mc)} \right]^2,
\qquad
R_8\equiv \bei\left[ \frac{137\mev}{\ms(\mc)+\md(\mc)} \right]^2
\ee
respectively. 
The scale $\mc=1.3\gev$ turns out to be convenient. $\ms(\mc)\approx
1.17\cdot \ms(2\gev)$.
The formula (\ref{ap}) exhibits very clearly the 
dominant uncertainties in 
$F_{\varepsilon'}$ which 
reside in the values of $\ms$, $\bsi$, $\bei$, $\Lms^{(4)}$ and $\OEE$.
Moreover, the partial 
cancellation between QCD penguin ($\bsi$) and electroweak 
penguin ($\bei$) contributions requires 
accurate 
values of $\bsi$ and $\bei$ for an acceptable estimate of $\epe$. 
Because of the accurate value $\mt(\mt)=165\pm 5~\gev$, the uncertainty 
in $\epe$ due to the top quark mass amounts only to a few percent. 
A very  accurate analytic formula 
for $F_{\varepsilon'}$ has been derived in \cite{buraslauten:93}
and its update can be found in \cite{EP99}.

Now, it has been known for some time that for central values of the 
input parameters the size of $\epe$
 in the Standard Model is well below the NA31 value of 
$(23.0\pm6.5)\cdot 10^{-4}$. Indeed, 
extensive NLO analyses with lattice and large--N estimates of 
$\bsi\approx 1$ and 
$\bei\approx 1$ performed 
first in \cite{BJLW,ROMA1} and after the top discovery in
\cite{ciuchini:95,BJL96a,ciuchini:96} have found $\epe$ in 
the ball park of 
$(3-7)\cdot 10^{-4}$ for $\ms(2~\gev)\approx 130~\mev$. 
 On the other hand 
it has been stressed repeatedly in \cite{AJBLH,BJL96a}  that 
for extreme values of $\bsi$, $\bei$ and $\ms$ still consistent with 
lattice, QCD sum rules and large--N estimates as well as 
sufficiently high values of $\IM\lambda_t$ and $\Lms^{(4)}$, 
a ratio $\epe$ as high as $(2-3)\cdot 10^{-3}$ could be obtained within 
the Standard Model. Yet, it has also been admitted that such simultaneously 
extreme values of all input parameters and consequently values of $\epe$
 close to the NA31 result 
are rather improbable in the Standard Model. 
Different conclusions have been reached in \cite{paschos:96}, where
values  $(1-2)\cdot 10^{-3}$ for $\epe$ can be found.
Also the Trieste group \cite{BERT98}, which calculated the parameters 
$\bsi$ and $\bei$
 in the chiral quark model, found 
$\epe=(1.7^{+1.4}_{-1.0})\cdot 10^{-3}$.
On the other hand using an  effective chiral
lagrangian approach, the authors in \cite{BELKOV} found $\epe$
consistent with zero.

\subsection{Hadronic Matrix Elements}
The main source of uncertainty in the calculation of
$\epe$ are the hadronic matrix elements $\langle Q_i(\mu)\rangle_I$.
They generally depend
on the renormalization scale $\mu$ and on the scheme used to
renormalize the operators $Q_i$. These two dependences are canceled by
those present in the Wilson coefficients $y_i(\mu)$ so that the
resulting physical $\epe$ does not (in principle) depend on $\mu$ and on the
renormalization scheme of the operators.  Unfortunately, the accuracy of
the present non-perturbative methods used to evalutate $\langle Q_i
\rangle_I$  is not
sufficient to have the $\mu$ and scheme dependences of
$\langle Q_i \rangle_I$ fully under control. 
We believe that this situation will change once the non-perturbative
calculations, in particular lattice calculations
improve.

As pointed out  in \cite{BJLW} the contributions of
$(V-A)\otimes(V-A)$ operators ($Q_i$ with i=1,2,3,4,9,10) 
to $\epe$ can be determined
from the leading CP conserving $K \to \pi\pi$ decays, for which the
experimental data is summarized in (\ref{eq:ReA0data}). 
The details of this approach will not be discussed here.
For the central value of $\IM\lambda_t$
these operators give a negative contribution to $\epe$ 
of about $-2.5\cdot 10^{-4}$. This shows that these
operators are only relevant if  $\epe$ is below $1 \cdot 10^{-3}$.
Unfortunately the matrix elements of the dominant $(V-A)\otimes(V+A)$
operators ($Q_6$ and $Q_8$) cannot be  determined by the 
CP conserving data and
one has to use  non-perturbative methods to estimate them.
Let us then briefly review the present status of
the corresponding non-perturbative parameters $\bsi$ and $\bei$ as well as 
of $\ms$, $\OEE$ and $\Lms^{(4)}$ which all enter the
evaluation of $\epe$ as seen in (\ref{ap}).

\subsection{The Status of $\ms$, $\bsi$, $\bei$, $\OEE$ and $\Lms^{(4)}$}
\subsubsection{$\ms$}
The values for $\ms(2\gev)$ extracted from quenched lattice calculations
and QCD sum rules before summer 1999 were
\begin{equation}\label{ms}
\ms(2\gev) =\left\{ \begin{array}{ll}
(110\pm20)\;\mev & {\rm (Lattice)}~\cite{GUPTA98,kenway98} \\
(124\pm22)\;\mev & {\rm (QCDS)}~\cite{QCDS} \end{array} \right.
\end{equation}
The value for QCD sum rules is an average \cite{EP99}
 over the results given in \cite{QCDS}.
QCD sum rules also allow to
derive lower bounds on the strange quark mass. It was found that generally
$\ms(2\gev)\gsim 100\,\mev$ \cite{MSBOUND}. 
The most recent  quenched lattice results:
$\ms(2\gev)= 106\pm 7~ \mev$ \cite{JLQCD99},
$\ms(2\gev)= 97\pm 4~ \mev$ \cite{ALPHA}, $\ms(2\gev)=105\pm 5~\mev$
\cite{GERMAN} are consistent with these bounds as well as (\ref{ms})
but have a smaller error. 
The unquenching seems to lower these values down
to $\ms(2\gev)= 84\pm 7~ \mev$ \cite{CPPACS}.
 We refer to the talks
of Ryan and Martinelli \cite{MARRY} for more details.
The most recent determination of $\ms$
from the hadronic $\tau$-spectral function \cite{PP} reads 
$\ms(2\gev)=(114\pm 23)\,\mev$ \cite{PP99} in a very good agreement
with (\ref{ms}) and lower that the value 
$\ms(2\gev)=(170^{+44}_{-55})\,\mev$ obtained by ALEPH
\cite{aleph:99} using this method early this year.
We conclude that the error on $\ms$ decreased considerably
during the last two years and
the central value is in the ball park of $\ms(2\gev)= 110\,\mev$
with smaller values coming from unquenched lattice QCD.

\subsubsection{$\bsi$ and $\bei$}
We recall that in the large--N limit $\bsi=\bei=1$ 
\cite{bardeen:87,Bardeen99}.
The values for $\bsi$ and $\bei$ obtained in various approaches
are collected in table~\ref{tab:317}. 
The lattice results have been obtained at $\mu=2\gev$.
The results in the large--N approach \cite{bardeen:87,Bardeen99} and 
the chiral quark model
correspond to scales below $1\gev$.
However,
 as a detailed numerical analysis in \cite{BJLW}
showed, $\bsi$ and $\bei$ are only weakly dependent on $\mu$.
Consequently
the  comparison
of these parameters
obtained in different approaches at different $\mu$ is meaningful.  
Next, the values coming from lattice and
chiral quark model are given in the NDR renormalization 
scheme. The
corresponding values in the HV scheme can be found
using approximate relations \cite{EP99} 
\be\label{NDRHV}
(\bsi)_{\rm HV}\approx 1.2 (\bsi)_{\rm NDR},
\qquad
 (\bei)_{\rm HV}\approx 1.2 (\bei)_{\rm NDR}.
\ee
The present results in the large-N approach are unfortunately
not sensitive to the renormalization scheme but this can be
improved \cite{Bardeen99,BP99}. 

Concerning the
lattice results for $B^{(1/2)}_{6}$,
the old results where in the range $0.6-1.2$
\cite{kilcup:91}. 
However,
a recent work in \cite{kilcup:99} shows
that lattice calculations of $\bsi$ are very uncertain 
and one has
to conclude that there are no solid predictions for
$B^{(1/2)}_{6}$ from the lattice at present. 
The average value of $\bsi$
in the large--N approach including full $p^2$ and 
$p^0/N$ contributions
and given in table~\ref{tab:317} is close to $1.0$
where the uncertainty comes from the variation of the
cut-off $\Lambda_c$ in the effective theory.
On the other hand, it
has been found \cite{Dortmund,Soldan} 
that a higher order term $\ord(p^2/N)$
enhances $\bsi$ to 1.5. This result is clearly interesting.
 Yet, in view of the fact that other $p^2/N$
terms as well as $p^4$ and $p^0/N^2$ terms have not been calculated,
it is premature to take this enhancement seriously. 
Finally, the chiral
quark model gives in the NDR scheme
the value for $\bsi$
as high as $1.33\pm 0.25$.

\begin{table}[thb]
\caption[]{ Results for $\bsi$ and $\bei$ obtained
in various approaches. 
\label{tab:317}}
\begin{center}
\begin{tabular}{|c|c|c|}\hline
  { Method}& $\bsi$& $B^{(3/2)}_8$  \\
 \hline
Lattice\cite{GKS,G67,APE}&$-$ &$0.69-1.06$  \\
Large$-$N\cite{DORT98,DORT99}& $0.72-1.10$ &$0.42-0.64$ \\
ChQM\cite{BERT98}& $1.07-1.58$ &$0.75-0.79$  \\
\hline
\end{tabular}
\end{center}
\end{table}

The status of $\bei$ looks better.
Most non-perturbative approaches 
find $\bei$ below unity. The suppression of $\bei$
below unity is rather modest (at most $20\%$) in the lattice 
approaches and in the chiral quark model. It is stronger in
the large--N  approach. Interestingly in the latter approach the ratio 
$\bsi/\bei\approx 1.72$ independently of the cut-off $\Lambda_c$.

Guided by the results presented above and biased to some
extent by the results from the large-N approach and lattice
calculations, we will use
in our numerical analysis below $\bsi$ and $\bei$ in
the ranges:
\be\label{bbb}
\bsi=1.0\pm0.3,
\qquad
\bei=0.8\pm 0.2
\ee
keeping always $\bsi\ge \bei$.

\subsubsection{$\OEE$ and $\Lms^{(4)}$}
The last estimates of $\OEE$ have been done more than ten years
ago \cite{donoghueetal:86}-\cite{lusignoli:89} 
and it is desirable to update these analyses which
 can be summarized by
$\OEE=0.25\pm 0.08~.$
In the numerical analysis presented below
we have incorporated the uncertainty in $\OEE$ by increasing
the error in $\bsi$ from $\pm 0.2$ to $\pm 0.3$.
Finally from $\alpha_s(\mz)=0.119\pm 0.003$ one deduces
$\Lms^{(4)}=(340\pm50)\mev$.
 
\subsection{Comments}
\subsubsection{General Remark}
We would like to emphasize that it would not be appropriate 
 to fit 
$\bsi$, $\bei$, $\ms$, $\Lms^{(4)}$, $\OEE$ and $\hat B_K$ in order to 
make the 
Standard Model compatible simultaneously with experimental values
on $\epe$, $\eps$ and the analysis of the unitarity triangle. Such an 
approach would be against the whole philosophy of searching for new physics 
with the help of loop induced transitions as represented by $\epe$ and 
$\eps$. Moreover it would not give us any clue whether the Standard
Model is consistent with the data on $\epe$. 
Indeed, it should be kept in mind that:

\begin{itemize}
\item
$\bsi$, $\bei$, $\hat B_K$ and $\OEE$ in spite of carrying the names of 
non-perturbative parameters, are really not 
parameters of the Standard Model as they can be calculated by means of 
non-perturbative methods in QCD. 
\item
$\ms$, $\Lms^{(4)}$, $\mt$, $\vcb$ and $|V_{ub}|$ are parameters of the 
Standard Model but there are better places than $\epe$ to 
determine them. In particular the usual determinations of these parameters 
can only marginally be affected by  physics beyond the Standard Model,
which is   not necessarily the case for $\eps$ and $\epe$.
\end{itemize}

Consequently, the only parameter to be fitted by direct CP violation is 
$\IM\lambda_t$. The 
numerical analysis of $\epe$ as a function of 
$\bsi$, $\bei$, $\ms$ and $\Lms^{(4)}$
presented below, should only give a global picture for 
which ranges of parameters the presence of new physics in $\epe$  
should be expected.

\subsubsection{The Issue of Final State Interactions}
In (\ref{eprime}) and (\ref{eq:epe2})
the strong phases $\delta_0\approx 37^\circ$ and
$\delta_2\approx -7^\circ$ are taken from experiment.
They can also be calculated from NLO chiral perturbation
for $\pi\pi$ scattering \cite{JGUM}. 
However, generally 
non-perturbative approaches to hadronic matrix elements are unable 
to reproduce them at present. 
As $\delta_I$ are factored out in (\ref{eprime}), in non-perturbative
calculations in which some final state interactions are
present in $\langle Q_i\rangle_I$ 
one should make the following
replacements in (\ref {eq:P12}) and (\ref{eq:P32}):
\be\label{FS0} 
\langle Q_i\rangle_I \to \frac{\RE\langle Q_i\rangle_I}
{(\cos\delta_I)_{\rm th}}
\ee
in order to avoid double counting of final state interaction
phases. Here $(\cos\delta_I)_{\rm th}$ is obtained in a given
non-perturbative calculation. 
Yet, in most calculations the phases are 
substantially smaller than found in experiment
\cite{bardeen:87,DORT99,DORT98,BERT98} and
$\langle Q_i\rangle_I \approx \RE\langle Q_i\rangle_I$.

The above point has been first discussed by the Trieste group
\cite{BERT98} who suggested
that in models in which at least the real part of $\langle Q_i\rangle_I$
can be calculated reliably, one should use
$(\cos\delta_I)_{\rm exp}$ in (\ref{FS0}).
As $(\cos\delta_0)_{\rm exp}\approx 0.8$ and 
$(\cos\delta_2)_{\rm exp}\approx 1$
this modification enhances $P^{(1/2)}$ by $25\%$ leaving $P^{(3/2)}$
unchanged. To our knowledge there is no method for hadronic matrix
elements which can provide $\delta_0\approx 37^\circ$  and consequently 
this suggestion may lead to an overestimate of the matrix 
elements and of $\epe$.

\subsubsection{$\epe$ and the $\Delta I=1/2$ Rule}
In  one of the first estimates of $\epe$, Gilman and Wise \cite{GW79}
used the suggestion of Vainshtein, Zakharov and Shifman \cite{PENGUIN}
that the amplitude $\RE A_0$ is dominated by the QCD-penguin operator
$Q_6$. Estimating $\langle Q_6 \rangle_0$ in this manner they predicted
a large value of $\epe$. Since then it has been understood 
\cite{DI12,kilcup:99,DORT99} that
as long as the scale $\mu$ is not much lower than $1~\gev$
the amplitude $\RE A_0$ is dominated by the current-current
operators $Q_1$ and
$Q_2$, rather than by $Q_6$. Indeed, at least in the HV
scheme the operator $Q_6$ does not contribute to $\RE A_0$
for $\mu=\mc$ at all, as its coefficient $z_6(\mc)$ relevant for
this amplitude vanishes. Also in the NDR scheme $z_6(\mc)$ is
negligible.

For decreasing $\mu$ the coefficient $z_6(\mu)$ increases and
the $Q_6$ contribution to $\RE A_0$ is larger.
However, if the analyses in \cite{DI12,kilcup:99,DORT99} are
taken into account, the operators $Q_1$ and $Q_2$ are responsible
for at least $90\%$ of $\RE A_0$ if the scale $\mu=1~\gev$ is considered.
Moreover, it should be stressed that whereas 
$Q_8$ operator is irrelevant for
the $\Delta I=1/2$ rule, it is important for $\epe$. 
Similarly, whereas the Wilson coefficients
$y_6(\mu)$ and $y_8(\mu)$ entering $\epe$ can be sensitive to
new physics as they receive contributions from very short distance
scales, this is not the case for $z_6(\mu)$, which due to GIM
mechanism receives contribution only from $\mu\le\mc$.
Therefore, even if $\langle Q_6 \rangle_0$ enters both
$\epe$ and $\RE A_0$, 
there is no strict relation between the large
value of $\epe$ and the $\Delta I=1/2$ rule as sometimes
stated in the literature. 
On the other hand if the long distance dynamics responsible
for the enhancement of  $\langle Q_{1,2} \rangle_0$ also
enhances $\langle Q_6 \rangle_0$, then some connection
between $\epe$ and the $\Delta I=1/2$ rule is possible.
There are some indications that this indeed could be
the case \cite{BERT98,Dortmund,Soldan}.

\subsubsection{$\epe$, $\bsi$, $\bei$ and $\ms$}
At this symposium there has been a vigorous discussion whether
the value of $\ms$ is really relevant for the estimate of $\epe$.
In non-perturbative approaches in which hadronic matrix elements
can only be calculated in terms of $\ms$, $F_\pi$, $F_K$ etc.,
it is obvious that the value of $\ms$ enters the estimate
of $\epe$. This is in particular the case of the large--N
approach. Moreover in this approach 
the values of $\bsi$ and $\bei$ are independent of $\ms$.

On the other hand,
as seen in (\ref{ap}), $\epe$ depends approximately on 
$\bsi$, $\bei$ and $\ms$
only through $R_6$ and $R_8$ defined in 
(\ref{RS}). If a given non-perturbative approach is able
to calculate directly the relevant hadronic matrix elements
and consequently $R_6$ and $R_8$ then in principle the
value of $\ms$ may not matter. If indeed $R_6$ and $R_8$
are independent of $\ms$ then through (\ref{RS}) there must be a 
quadratic dependence of $\bsi$ and $\bei$ on $\ms$. As stated
above this dependence is not observed in the large--N approach
\cite{bardeen:87,DORT98,DORT99}. As discussed by Martinelli 
 at this symposium, this question is being investigated  
in the lattice approach at present. 
On the other hand
there are results in the literature showing a strong 
$m_s$-dependence of the $B_i$ parameters. This is the case
for $\bsi$ in the chiral quark model where $\bsi$ scales
like $\ms$ \cite{BERT98}. Similarly values for $B_7^{(3/2)}$
calculated in \cite{DER1} show a strong $\ms$-dependence.
However, $\bei$ calculated in the chiral quark model
shows a very weak dependence on $\ms$.
This discussion shows that comparision of $(\bsi,\bei)$
obtained in various approaches has to be done with care.

Personally, I think that the value of $\ms$ is relevant for
the matrix elements $\langle Q_6\rangle_0$ and 
$\langle Q_8\rangle_2$ and
$(\bsi,\bei)$ are nearly $\ms$ independent. As both operators
(after Fierz transformation)
have the density-density structure, their matrix elements
are proportional to the square of the quark condensate and
hence proportional to $1/\ms^2$.
In this context it should be recalled that the anomalous
dimenions of $Q_6$ and $Q_8$ are in a good approximation
equal twice the anomalous dimension of the mass operator.
As a result of this, the products 
$y_6(\mu)\langle Q_6(\mu) \rangle_0$ and
$y_8(\mu)\langle Q_8(\mu) \rangle_2$ and the corresponding
contributions to $\epe$ are only very weakly $\mu$-dependent.

\subsection{Numerical Estimates of $\epe$}
\subsubsection{Munich Analysis}
We will begin by presenting the analysis in \cite{EP99}.
In table~\ref{tab:inputp} we summarize the input parameters
used there.
The value 
of $\ms(\mc)$,
corresponds roughly to $\ms(2~\gev)=(110\pm20)\mev$ as obtained 
in lattice simulations. $\IM\lambda_t$ is given in (\ref{imt})
except that in evaluating $\epe$ the correlation in $\mt$
between $\IM\lambda_t$ and $F_{\varepsilon'}$ has to be taken
into account.
In what follows we will present the results of 
two types of  numerical analyses of
$\epe$ which use Monte Carlo and scanning methods 
discussed already in section 2.

\begin{table}[thb]
\caption[]{Collection of input parameters.
$\ms(\mc)=1.17\ms(2\gev)$.
\label{tab:inputp}}
\vspace{0.4cm}
\begin{center}
\begin{tabular}{|c|c|c|c|}
\hline
{\bf Quantity} & {\bf Central} & {\bf Error} & {\bf Reference} \\
\hline
$\Lms^{(4)}$ & $340 \mev$ & $\pm 50\mev$ & \cite{PDG,BETKE} \\
$\ms(\mc)$ & $130\mev$    & $\pm 25\mev$ & See Text\\
$\bsi $ & 1.0 & $\pm 0.3$ & See Text\\
$\bei $ & 0.8 & $\pm 0.2$ & See Text\\
\hline
\end{tabular}
\end{center}
\end{table}

\begin{figure}
\begin{center}
% GNUPLOT: LaTeX picture with Postscript
\setlength{\unitlength}{0.1bp}
\begin{picture}(3492,2160)(0,0)
\special{psfile=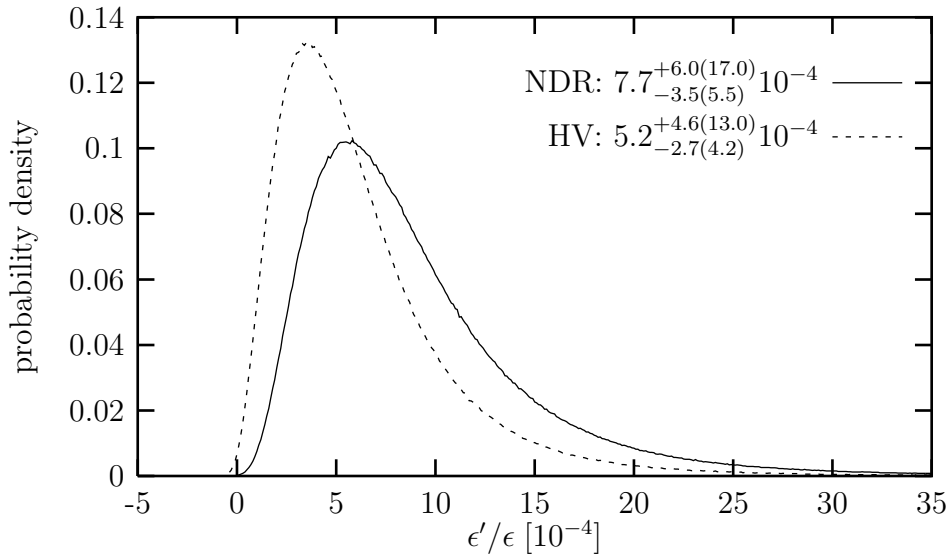 llx=0 lly=0 urx=698 ury=504 rwi=6980}
\put(2968,1609){\makebox(0,0)[r]{HV: $5.2^{+4.6(13.0)}_{-2.7(4.2)} 10^{-4}$}}
\put(2968,1813){\makebox(0,0)[r]{NDR: $7.7^{+6.0(17.0)}_{-3.5(5.5)} 10^{-4}$}}
\put(1896,100){\makebox(0,0){$\epsilon'/\epsilon \;[10^{-4}]$}}
\put(0,1195){%
\special{ps: gsave currentpoint currentpoint translate
270 rotate neg exch neg exch translate}%
\makebox(0,0)[b]{\shortstack{probability density}}%
\special{ps: currentpoint grestore moveto}%
}
\put(3392,230){\makebox(0,0){35}}
\put(3018,230){\makebox(0,0){30}}
\put(2644,230){\makebox(0,0){25}}
\put(2270,230){\makebox(0,0){20}}
\put(1896,230){\makebox(0,0){15}}
\put(1522,230){\makebox(0,0){10}}
\put(1148,230){\makebox(0,0){5}}
\put(774,230){\makebox(0,0){0}}
\put(400,230){\makebox(0,0){-5}}
\put(350,2060){\makebox(0,0)[r]{0.14}}
\put(350,1813){\makebox(0,0)[r]{0.12}}
\put(350,1566){\makebox(0,0)[r]{0.1}}
\put(350,1319){\makebox(0,0)[r]{0.08}}
\put(350,1071){\makebox(0,0)[r]{0.06}}
\put(350,824){\makebox(0,0)[r]{0.04}}
\put(350,577){\makebox(0,0)[r]{0.02}}
\put(350,330){\makebox(0,0)[r]{0}}
\end{picture}
\end{center}
\vspace{-6mm}
\caption{Probability density distributions for $\epe$ in NDR and HV
schemes \cite{EP99}.}
\label{g1}
\end{figure}

Using the first method we find the probability density 
distributions for
$\epe$ in fig.~\ref{g1}.
From this distribution
we deduce the following results:
\begin{equation}\label{eq:eperangefinal}
\epe =\left\{ \begin{array}{ll}
( 7.7^{~+6.0}_{~-3.5}) \cdot 10^{-4} & {\rm (NDR)} \\
( 5.2^{~+4.6}_{~-2.7}) \cdot 10^{-4} & {\rm (HV)} \end{array} \right.
\end{equation}
The difference between these two results indicates the left over
renormalization scheme dependence.
Since, the resulting probability density distributions for
$\epsilon'/\epsilon$ are very asymmetric with  very long
tails towards large values we quote the medians
and the $68\%(95\%)$ confidence level intervals. 
Using the second method we find :
\begin{equation}
~~~~~1.05 \cdot 10^{-4} \le \epe \le 28.8 \cdot 10^{-4}\qquad {\rm (NDR)}.
\label{eq:eperangenew}
\end{equation}
and
\begin{equation}
~~~~~0.26 \cdot 10^{-4} \le \epe \le 22.0 \cdot 10^{-4}\qquad {\rm (HV)}.
\label{hv:eperangenew}
\end{equation}

We observe that
$\epe$ is generally lower in the HV scheme if the same values for
$B_6^{(1/2)}$ and $B_8^{(3/2)}$ are used in both schemes. 
Since the present non-perturbative methods do not have renormalization
scheme dependence fully under control we think that such a treatment of
$B_6^{(1/2)}$ and $B_8^{(3/2)}$ is the proper way of estimating
scheme dependences at present.
Assuming, on the other hand, that the values in (\ref{bbb}) correspond
to the NDR scheme and using the relation (\ref{NDRHV}), we find for
the HV scheme the range 
$0.58 \cdot 10^{-4} \le \epe \le 26.9 \cdot 10^{-4}$ which is
much closer to the NDR result in (\ref{eq:eperangenew}). 
This exercise shows that it is very desirable to have the
scheme dependence under control.

We observe that the most probable values for $\epe$ in the NDR scheme
 are in the
ball park of $1 \cdot 10^{-3}$. They are lower by roughly $30\%$ in the
HV scheme if the same values for $(\bsi,\bei)$ are used.
On the other hand the ranges in (\ref{eq:eperangenew}) and
(\ref{hv:eperangenew}) show that for
particular choices of the input parameters, values for $\epe$ as high as
$(2-3)\cdot 10^{-3}$ cannot be excluded at present. Let us study 
this in  more detail.

In table~ \ref{tab:31731} we show the values 
 of $\epe$ in units of $10^{-4}$ 
for specific values of $\bsi$, $\bei$ and $\ms(\mc)$ as calculated
in the NDR scheme. The corresponding values in the HV scheme
are lower as discussed above.
The fourth column shows the results for central values of all remaining
parameters. The comparison of the the fourth and the fifth column
 demonstrates
how $\epe$ is increased when $\Lms^{(4)}$ is raised from $340~\mev$
to $390~\mev$. As stated in (\ref{ap}) $\epe$ is roughly proportional
to $\Lms^{(4)}$. Finally, in the last column maximal values of $\epe$
are given.
To this end we have scanned all parameters relevant for
the analysis of $\IM\lambda_t$ within one
standard deviation and have chosen the highest value of
$\Lms^{(4)}=390\mev$. Comparison of the last two columns demonstrates
the impact of the increase of $\IM\lambda_t$ from its
central to its maximal value and of the variation of $\mt$.

Table~\ref{tab:31731} gives a good insight in the dependence of
$\epe$ on various parameters which is roughly described
by (\ref{ap}). We observe the following hierarchies:

\begin{itemize}
\item
The largest uncertainties reside in $\ms$, $\bsi$ and $\bei$.
\item
The combined uncertainty due to $\IM\lambda_t$ and $\mt$,
present both in $\IM\lambda_t$ and $F_{\varepsilon'}$,
is approximately $\pm 25\%$. The uncertainty due to
$\mt$ alone is only $\pm 5\%$.
\item
The uncertainty due to $\Lms^{(4)}$ is approximately $\pm 16\%$.
\item
The uncertainty due to $\OEE$ is approximately $\pm 12\%$.
\end{itemize}

The large sensitivity of $\epe$ to $\ms$ has been known 
since the analyses in the eighties. In the context of
the KTeV result this issue has been first 
analyzed in \cite{Nierste}.
It has been found that provided $2\bsi-\bei\le 2$ the
consistency of the Standard Model with the KTeV result
requires the $2\sigma$ bound $\ms(2\gev)\le 110\mev$.
Our analysis is compatible with these findings.

\begin{table}[thb]
\caption[]{ Values of $\epe$ in units of $10^{-4}$ 
for specific values of $\bsi$, $\bei$ and $\ms(\mc)$
and other parameters
as explained in the text \cite{EP99}.
\label{tab:31731}}
\begin{center}
\begin{tabular}{|c|c|c|c|c|c|}\hline
 $B^{(1/2)}_6$& $B^{(3/2)}_8$ & $\ms(\mc)[\mev]$ &
  Central & $\Lms^{(4)}=390\mev $ & Maximal \\ \hline
      &     & $105$ &  20.2 & 23.3 & 28.8\\
 $1.3$&$0.6$& $130$ &  12.8 & 14.8 & 18.3\\
      &     & $155$ &   8.5 &  9.9 & 12.3 \\
 \hline
      &     & $105$ &  18.1 & 20.8 & 26.0\\
 $1.3$&$0.8$ & $130$ & 11.3 & 13.1 & 16.4\\
      &     & $155$ &   7.5 &  8.7 & 10.9\\
 \hline
      &     & $105$ &   15.9 & 18.3 & 23.2\\
 $1.3$&$1.0$ & $130$ &  9.9 &  11.5 & 14.5\\
      &     & $155$ &   6.5  &  7.6 &  9.6\\
 \hline\hline
      &     & $105$ &   13.7 & 15.8 & 19.7\\
 $1.0$&$0.6$ & $130$ &  8.4 &  9.8& 12.2 \\
      &     & $155$ &   5.4 &  6.4 & 7.9 \\
 \hline
      &     & $105$ &   11.5 & 13.3 & 16.9\\
$1.0$&$0.8$ & $130$  &  7.0 &   8.1 & 10.4\\
     &     & $155$ &   4.4 &    5.2 &  6.6\\
 \hline
     &     & $105$ &   9.4 &   10.9 & 14.1 \\
$1.0$&$1.0$ & $130$  &  5.5 &   6.5 &  8.5 \\
     &     & $155$ &   3.3 &    4.0 &  5.2\\
 \hline
\end{tabular}
\end{center}
\end{table}

\subsubsection{Comparision with Other Analyses}
In table \ref{tab:31738} we compare the results discussed
above with the most recent
results obtained by other groups. 
Details of these analyses can be found in other contributions
to this symposium and in references quoted in the table.
There exists no recent 
phenomenological analysis from the Trieste group \cite{BERT98}
and we quote their 1998 result. The values for $\bsi$ and
$\bei$ are given in the NDR scheme. The corresponding
values in the HV scheme can be found by using (\ref{NDRHV}).
All groups use the Wilson coefficients calculated in
\cite{NLOP}--\cite{ROMA1} and the differences in $\epe$
result dominantly from different values of $\bsi$,
$\bei$ and $\ms$ and to some extend in different values
in input parameters (like $B_K$) needed for the determination
of $\IM\lambda_t$. The results for $\epe$ are given in the
NDR scheme except for the result from Trieste which corresponds
to the HV scheme. The values of $\ms(\mc)$ used by all groups
are very close to the ones given in table~\ref{tab:inputp}. 

We observe that the results from Munich, Rome and Dortmund are
compatible with each other and generally well below the
experimental data. The values of $\epe$ for central values
of parameters obtained by the Dortmund group are in good agreement 
with those of the
Munich group. On the other hand as discussed above and also found
by the Dortmund group for extreme choices of input parameters
values for $\epe$ consistent with experiment can be obtained.
In the framework of an effective chiral
lagrangian approach \cite{BEL}
$B_6^{(1/2)}$ and $B_8^{(3/2)}$ cannot be calculated.
For $\bsi=\bei=1$
values for $\epe$  consistent with zero are obtained.

The Trieste group finds generally higher values of
$\epe$, with the central value around $17\cdot 10^{-4}$ and consequently
consistent with the experimental findings. 
Similarly the Dortmund group finds the central values for
$\epe$ in the ball park of $15\cdot 10^{-4}$ if they use
$\bsi\approx 1.5$ as obtained by including one $p^2/N$ term.

\begin{table}[thb]
\caption[]{ Results for $\epe$ in units of $10^{-4}$ obtained
by various groups. The labels (MC) and (S) in the last column
stand for ``Monte Carlo'' and ``Scanning'' respectively, as discussed
in the text. 
\label{tab:31738}}
\begin{center}
\begin{tabular}{|c|c|c||c|}\hline
  {\bf Reference}& $B^{(1/2)}_6$& $B^{(3/2)}_8$ & 
 $\epe[10^{-4}]$ \\ \hline
Munich
\cite{EP99}& $1.0\pm 0.3$ &$0.8\pm0.2$ &  
$7.7^{+6.0}_{-3.5}$ (MC) \\
Munich
\cite{EP99}& $1.0\pm 0.3$ &$0.8\pm0.2$ & $1.1\to 28.8$ (S) \\
\hline
Rome
\cite{ROMA99}& $0.83\pm 0.83$ &$0.71\pm0.13$ &  
$4.7^{+6.7}_{-5.9}$ (MC) \\
\hline
Dortmund
\cite{Dortmund}& $0.72-1.10$ &$\bsi/1.72$ &  
$2.1\to 26.4$ (S) \\
\hline
Trieste
\cite{BERT98}& $1.33\pm 0.25$ &$0.77\pm0.02$ &  
$7\to 31$ (S) \\
\hline
Dubna-DESY
\cite{BEL} 
& $1.0$ &$1.0$ & $-3.2 \to 3.3$ (S) \\
\hline
\end{tabular}
\end{center}
\end{table}

\section{$\epe$ Beyond the Standard Model}
As we have seen the estimates of $\epe$ within the Standard Model (SM) 
are generally below
the data but in view of large theoretical uncertainties stemming
from hadronic matrix elements one cannot firmly conclude that the
data on $\epe$ indicate new physics.
However there is
still a lot of 
room for non-standard contributions to $\epe$ and the apparent
discrepancy between the SM estimates and the data invites
for speculations. Indeed  results from NA31, KTeV and NA48
prompted several  analyses
of $\epe$ within various extensions of the Standard Model
like general supersymmetric models \cite{Nierste,MM99,BS99},
models with anomalous gauge couplings \cite{HE} and 
models with additional
fermions and gauge bosons \cite{Frampton}.
Unfortunately several of these extensions have
many free parameters and are not very conclusive.

On the other hand it is  clear that the $\epe$ data puts models 
in which there are new positive contributions to $\eps$ and 
negative contibutions to 
$\varepsilon'$ in serious difficulties. In particular 
as analyzed in \cite{EP99} the two Higgs Doublet Model II
can either be ruled out  with improved 
hadronic matrix elements or a  
powerful lower bound on $\tan\beta$ can 
be obtained from $\epe$.
In the Minimal Supersymmetric Standard Model, in addition to charged
Higgs exchanges in loop diagrams, also charginos contribute.
For suitable  choice of the
supersymmetric parameters, the chargino contribution
can enhance $\epe$  with respect to the Standard
Model expectations \cite{GG95}.
Yet, generally
the most conspicuous effect of
minimal supersymmetry is a depletion of $\epe$.
The situation can be different in more general
models in which there are more parameters than
in the two Higgs doublet model II and in the MSSM, in particular
new CP violating phases.
As an example, in general supersymmetric models
$\epe$ can be made consistent with experimental
findings through the contributions of the chromomagnetic
penguins \cite{GMS,Nierste,MM99,BS99} and enhanced $Z^0$-penguins
with the opposite sign to the one in the Standard Model
\cite{ISI,BS98,BS99}.

While substantial new physics contributions to $\epe$ from
chromomagnetic penguins and modified Z-penguins appear
rather plausible, one can give rather solid arguments 
that new physics should have only a minor impact on the
QCD-penguins represented by the operator $Q_6$. The point is
that the contribution of $Q_6$ to $\epe$ can  generally
be written as follows
\begin{equation}\label{QCD}
\left(\epe\right)_{Q_6}\approx \tilde R_6 \left[11-1.3 E(\mt,..)\right]
\end{equation}
where
\begin{equation}\label{TR6}
\tilde R_6=\IM\lambda_t
\left[\frac{110\mev}{\ms(2~\gev)}\right]^2
\left(\frac{\Lms^{(4)}}{340~\mev}\right) \bsi
\ee
and $E(\mt,...)$ results from QCD penguin diagrams in the full
theory which in addition to $W$-boson and top-quark exchanges
may receive contributions from new particles. As in the
Standard Model $E(\mt)\approx 0.3$, the contribution of $Q_6$
to $\epe$ is dominated by ``11" which results from the operator
mixing between $Q_6$ and other operators, in particular 
the current-current operator $Q_2$. The latter mixing taking
place at scales below $\mw$ is unaffected by new physics
contributions which can enter (\ref{QCD}) only through the
function $E$. One would need an order of magnitude
enhancement of $E$ through new physics contributions
in order to see a $40\%$ effect in the contribution of $Q_6$
to $\epe$ and unless the sign of $E$ is reversed one would
find rather a suppression of $\epe$ with respect to the SM
expectations than a required enhancement. This discussion
shows that a substantial enhancement of QCD-penguins through
new physics contribution as suggested recently in \cite{Chanowitz}
is rather implausible. On the other hand in the case of the
operator $Q_8$ one has instead of (\ref{QCD})
\begin{equation}\label{EW}
\left(\epe\right)_{Q_8}\approx \tilde R_8 \left[-10\right] Z(\mt,..)
\end{equation}
where $\tilde R_8$ is given by (\ref{TR6}) with $\bsi$ replaced
by $\bei$ and $Z(\mt,...)$ results from 
electroweak penguin diagrams, in
particular Z-penguin diagrams. Again ``$-10$" comes from
operator mixing under renormalization and new physics can enter
(\ref{EW}) only through the function $Z$. But this time any
sizable impact of new physics on the function $Z$ translates
directly into a sizable impact on $\epe$.

\section{Summary}
 As we have seen,
the estimates of $\epe$ in the Standard Model are typically below 
the experimental data. However,
as the scanning analyses show,
for suitably chosen parameters, $\epe$ in the Standard 
Model can be made consistent with data. Yet, this happens only if all 
relevant parameters are simultaneously close to their extreme values. 
This is clearly seen in table~\ref{tab:31731}.
Moreover, the probability density distributions for $\epe$ in
fig.~\ref{g1} indicates that values of $\epe$ in the ball park of
the experimental grand average $21.2 \cdot 10^{-4}$
are rather improbable.

In spite of a  possible ``disagreement" of the Standard Model
with the data, one should realize that certain
features present in the Standard Model are confirmed by
the experimental results. 
Indeed, the sign and the order of magnitude of $\epe$ 
predicted by the SM turn out to agree with the data.
In obtaining these results renormalization group evolution
between scales $\ord(\mw)$ and $\ord(1\gev)$, an important
ingredient in the evaluation of the Wilson coefficients $y_i(\mu)$,
 plays a crucial role. As analyzed in \cite{BU} without these 
renormalization group effects 
$y_6$ and $y_8$ would be tiny and $\epe$ at most $\ord(10^{-5})$
in vast disagreement with the data. Since these effects are present
in all extensions of SM, we conclude that we are  probably
on the right track.

Unfortunately, in view of very large hadronic and substantial parametric 
uncertainties, it 
is impossible to conclude at present whether new physics contributions are 
indeed required to fit the data.
Yet as we stressed above, there is
still a lot of 
room for non-standard contributions to $\epe$.
The most plausible sizable new contributions
could come from chromomagnetic penguins in general
supersymmetric models and modified Z-penguins.
On the other hand substantial modification of QCD-penguins
through new physics are rather implausible.

In view of large hadronic uncertainties it is difficult to conclude what 
is precisely the impact of 
the $\epe$-data on the CKM matrix. However, as analyzed in \cite{EP99}
there are indications  
that the lower limit on 
$\IM\lambda_t$ is improved. The same applies to the lower limits for the 
branching ratios for $K_L\to\pi^0\nu\bar\nu$ and 
$K_L\to\pi^0 e^+ e^-$ decays.

The future of $\epe$ in the Standard Model and in its extensions depends on 
the progress in the reduction of parametric and in particular 
hadronic uncertainties. 
In this context it is essential to get full control over
renormalization scheme and renormalization scale dependences
of hadronic matrix elements.
Personally I believe that the only
hope for making Standard Model naturally consistent with the data,
without streching all parameters to their extreme values, is 
the value of $\ms(2~\gev)$ in the ball park of $90-100 \mev$,
$\bsi$ in the ball park of $1.5-2.0$ and $\bei<1.0$ (both in
the NDR scheme).
The required values for $\ms$ seem to be found in 
lattice calculations but the story is not finished yet.
Values of $\bsi$ in the ball park of $1.5-2.0$ are suggested
by the chiral quark model calculations \cite{BERT98} and a
higher order term $\ord(p^2/N)$ in the large--N approach
\cite{Dortmund}. Yet, from my point of view, it is not clear
how well the chiral quark model approximates QCD and whether other
higher order terms in the large--N approach will weaken the
indicated enhancement of $\bsi$.
We should hope that the new efforts by the
lattice community will help in clarifying the situation.
An interesting work in this direction in the framework
of the large--N approach can also be found in \cite{BP99}.
In any case $\epe$ already played a decisive role in establishing direct 
CP violation in nature and its rather large value gives additional strong 
motivation for searching for this phenomenon in other decays. 

{\bf Acknowledgements}\\
I would like to thank Jonathan Rosner and Bruce Winstein
for inviting me to such an exciting symposium. I would
also like to thank all my collaborators for most
enyojable time we had together and the authors
of \cite{ROMA99,BERT98,Dortmund,BEL} for informative
discussions.

This work has been supported by the German Bundesministerium f\"ur 
Bildung und Forschung
under contract 06 TM 874. Travel support from Max-Planck Institute
for Physics in Munich is gratefully acknowledged.

\vfill\eject

\end{document}